# Sub-100 Hz Intrinsic Linewidth 852 nm Silicon Nitride External Cavity Laser

Hani Nejadriahi,[1,*] Eric Kittlaus,[1] Debapam Bose,[2] Nitesh Chauhan,[2] Jiawei Wang,[2] Mathieu Fradet[1], Mahmood Bagheri[1], Andrei Isichenko[2], David Heim[2], Siamak Forouhar[1], and Daniel J. Blumenthal[2]

[1]*Jet Propulsion Laboratory, California Institute of Technology, Pasadena, CA 91001, USA*
[2]*Department of Electrical and Computer Engineering, University of California, Santa Barbara, Santa Barbara, CA, 93106, USA*
*hani.neajadriahi@jpl.nasa.gov

**We demonstrate an external cavity laser with intrinsic linewidth below 100 Hz around an operating wavelength of 852 nm, selected for its relevance to laser cooling and manipulation of cesium atoms. This system achieves a maximum CW output power of 24 mW, wavelength tunability over 15 nm, and a side-mode suppression ratio exceeding 50 dB. This performance level is facilitated by careful design of a low-loss integrated silicon nitride photonic circuit serving as the external cavity combined with commercially available semiconductor gain chips. This approach demonstrates the feasibility of compact integrated lasers with sub-kHz linewidth centering on the needs of emerging sensor concepts based on ultracold atoms and can be further extended to shorter wavelengths via selection of suitable semiconductor gain media.**

Emerging instrument concepts based on harnessing ultracold atoms present unique opportunities across a spectrum of sensing and metrology applications [1,2], ranging from atom interferometer inertial sensors [2,3], coherent communications [4], atomic clocks [5,6], and frequency synthesizers [7,8]. Common to all of these instruments is the requirement for sophisticated laser optics systems (LOS) to manipulate and interrogate the atomic species under test. [9-12]. Despite significant progress in miniaturizing reference cells and microwave interfaces, many cold atom sensors require substantial supporting laboratory hardware, posing ongoing challenges toward achieving compact and reliable field-deployable instruments [12-15].

High-performance, frequency-stable lasers are essential components of such systems. Stringent requirements on laser noise level and output power, combined with the need for operation at specific wavelengths relevant to each atomic species, necessitate the development of application-specific laser oscillators for field deployment. Lasers based on semiconductor gain media are one attractive option, offering compact size, low power consumption, and the capability to be engineered for operation across a range of relevant wavelengths. However, existing commercial semiconductor lasers at sub-micron wavelengths often have modest noise performance that is insufficient for all atomic sensing applications. For example, distributed Bragg reflector (DBR) lasers represent one of the preferred schemes for stable operation, but short cavity lengths typically limit intrinsic laser linewidth to >10 kHz [13,15].

Recent advances in integrated photonics have resulted in significant progress towards developing low-noise lasers for visible and near-IR (NIR) applications, with various approaches aimed at reducing fundamental linewidth. Notable examples include silicon nitride stimulated Brillouin lasers (SBL), achieving intrinsic linewidths of 24 Hz at 780 nm [16] and 296 Hz at 674 nm [17]. Additionally, silicon nitride visible light injection-locked lasers (SIL) have demonstrated fundamental linewidths as low as 600 mHz (sub-Hz) at 780 nm [18], offering an important integrated solution for visible and NIR applications. Besides common approaches such as frequency doubling of telecom lasers [19], another common design is the extended cavity laser (ECL), widely used in atomic experiments, which traditionally relies on mechanical components. While integrated ECLs with low linewidths have been successfully implemented at telecommunications wavelengths [20,21], progress at visible and NIR atomic wavelengths has been limited. Since atomic transitions typically occur in the visible to near-IR range, developing on-chip laser sources at these wavelengths is crucial for future advancements in atomic sensing and metrology applications. [22].

In this work, we demonstrate an ECL with sub-100 Hz intrinsic linewidth operating at 852 nm, fabricated in the ultra-low loss silicon nitride platform designed for sub-micron wavelengths. The laser employs a configuration similar to that reported in [23] with an intra-cavity consisting of a Vernier 2-ring filter and tunable Sagnac loop mirror is engineered to provide feedback to a reflective semiconductor optical amplifier (RSOA). We target operation around the Cesium $D_2$ line (852 nm) for applications to atom-based sensors such as atom interferometer gravimeters currently under investigation for NASA applications [11]. This work can be extended to other wavelengths based on the choice of semiconductor gain medium. We report high side mode suppression ratio (> 50 dB), and high output power (~ 24 mW), with an intrinsic linewidth of 65 Hz. These characteristics are attractive for an array of precision metrology and sensing applications at sub-micron wavelengths. [1-3].

**Results.** To achieve single-mode lasing with low phase noise at 852 nm, the laser needs a long intracavity photon lifetime which is achieved by increasing the cavity Q and lowering the propagation losses at this wavelength. Additionally, the intracavity filter must have high spectral resolution to ensure single-mode oscillation [24-27]. To meet these requirements at the desired operation wavelength, we utilize the low loss silicon nitride ($Si_3N_4$) integrated photonic platform that achieves low loss and high Q at visible and

NIR wavelengths [27-29]. These low loss designs employ a thin silicon nitride layer [30-34] and a high-purity top oxide cladding to reduce the overlap of the optical mode with the nitride core sidewalls and hence the scattering induced losses. The high-purity oxide is realized through a combination of low-pressure chemical vapor deposition (LPCVD) and plasma-enhanced chemical vapor deposition (PECVD) growth techniques combined with a high-temperature anneal at 1050°C for over 12 hours. The resulting device cross-section is shown in Fig. 1(a). [30]. The external cavity devices fabricated here exhibited losses around 10 dB/m at 852 nm. The low propagation losses enabled by the $Si_3N_4$ platform allow for the design of a relatively large on-chip external cavity. Building on these results, we developed a dual-chip ECL design comprising of an input taper, a narrowband filter and mirror in a $Si_3N_4$ based photonic chip, and a GaAs RSOA chip is illustrated in Fig. 1(b & c). Similar designs have been employed to demonstrate lasing at telecom wavelengths (~1550 nm) [22,24,25]. We selected this configuration because it enables the extension of the cavity length and incorporates narrowband spectral filtering through Vernier ring resonators [20], facilitating intracavity resonant excitation with reasonable output power and high side-mode suppression ratio. The use of an end mirror in the form of a Sagnac loop allows the light to traverse the rings twice per round trip, extending the optical path and reducing the linewidth. [22,35,36].

The ECL's design comprises four major sections: the gain chip, input taper, Vernier ring filter, and loop mirror output coupler. The gain medium is a RSOA (Photodigm, Inc. 850SAF100) which is butt-coupled to the transverse electric mode of the external cavity. The RSOA chip has an optical length of 1 mm and center wavelength between 845-855 nm. The back facet is high-reflection coated (R ~ 90%) to provide double-pass amplification. To avoid unwanted back reflections into the amplifier, the front facet of the amplifier has an anti-reflection coating with a reflectivity of 0.1% and the waveguide is angled at 5 degrees with respect to the normal. The beam waist is approximately 14 μm × 3 μm in diameter at the RSOA output. A spot size converter efficiently transfers light from the gain chip to the silicon nitride chip, as shown in Fig. 1(c). The taper extends up to 11.1 μm, with typical ~50% coupling ratio measured across multiple chips. This tapered waveguide is angled at 10.5 degrees to match the output of the angled-facet RSOA chip and further minimize reflections.

Following the tapered waveguide, two ring resonators with slightly different radii (and corresponding different free spectral ranges) using 3 μm wide waveguides and 300 nm ring-bus waveguide gap are cascaded in a Vernier configuration to serve as a narrowband spectral filter inside the cavity. The individual spectra of the two cascaded add-drop ring resonators and the corresponding Vernier envelope are similar to previously designed Vernier rings as shown in [21,25,26]. The free spectral range (FSR) of the Vernier envelope thus can be written as $FSR_{Vernier} = \lambda^2/2\pi n_g|\Delta R|$. To ensure that exactly one Vernier mode lies within 15 nm 3-dB optical bandwidth of the gain chip, we chose two rings with radii of 2104.3 μm and 2100 μm respectively, resulting in FSRs of 31.26 pm and 31.19 pm. This offset in the FSR is designed to be significant enough to prevent low round-trip losses in adjacent peaks but sufficiently small so that the next Vernier overlap occurs at 15.6 nm, which sits outside of the RSOA chip's gain bandwidth. The individual ring resonators are equipped with resistive heaters (in this case Ti/Au) for thermal tuning and wavelength selectivity.

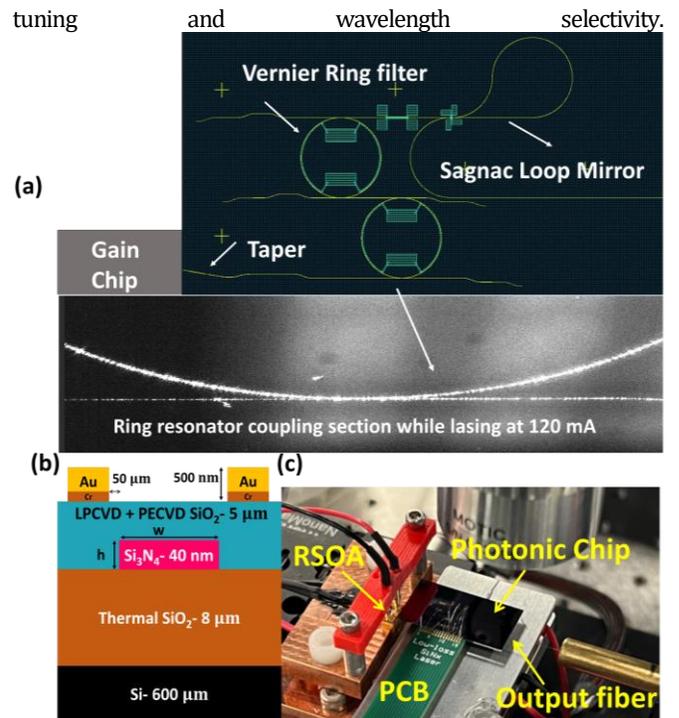

**Figure 1:** (a) Schematic of the external laser cavity design, comprising an input taper region, dual ring resonators to create the Vernier filter, and a Sagnac loop mirror – inset shows the optical microscope image of the bus-waveguide coupling section of the ring resonator while lasing(b) Cross-sectional view of the waveguide geometry; w = 3 μm was used for the Vernier ring filters, while the waveguide is tapered down to w = 1.25 μm for the Sagnac loop mirror. (c) Photograph of the characterization setup used for the alignment of the RSOA gain chip on a 5-axis stage to the $Si_3N_4$ chip (wire-bonded to a PCB) and lensed fiber at the output for spectral measurements.

Finally, to serve as the output coupler and mediate optical feedback, we incorporate a Sagnac loop mirror. Adjusting the length and gap between adjacent waveguides allows for tuning of the mirror's reflectivity. In our design, we utilize single-mode waveguides measuring 1.25 μm in width and 100 μm in length, with a 400 nm gap. Single mode waveguides were chosen for the directional coupler part of the mirror to avoid potential modal crosstalk. Through simulation and experimental validation, we determine that this configuration yields a coupling ratio of 20-30%, corresponding to a mirror reflectivity of 60-80%. The highly reflective end of the gain chip, in conjunction with the loop mirror, satisfies the Fabry-Perot resonance condition.

The experimental setup for the ECL involves an RSOA chip on a 5-axis stage that is butt-coupled to a $Si_3N_4$ chip as shown in Fig 1(c). Pump current and heater control for tuning are provided by probes connected to a voltage/current source, with a thermo-electrical cooler (TEC) maintaining the RSOA at a constant 25°C. The $Si_3N_4$ chip is also temperature-controlled by a TEC, keeping it between 25-27°C. Light from the $Si_3N_4$ chip is collected using a lensed tapered fiber for optical spectrum measurements and on-chip power measurements are performed using a free-space lens and photodetector directly at the waveguide output.

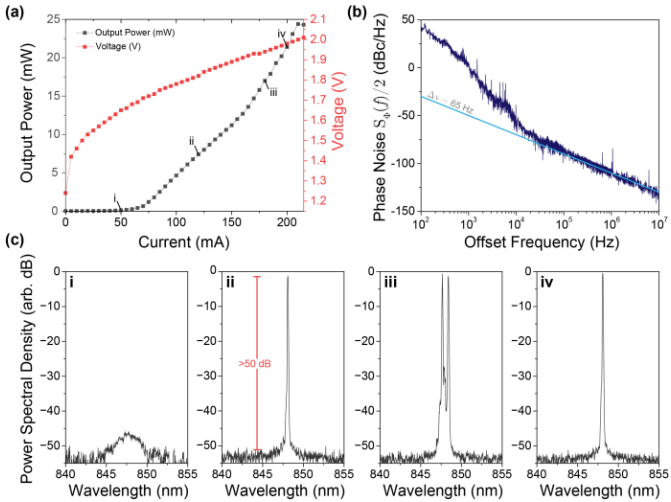

**Figure 2:** (a) Free-space power at the output of the silicon nitride chip plotted as a function of pump current and corresponding I-V curve. (b) Optical phase noise measurement as characterized with a delayed self-heterodyne frequency discriminator. (c) Optical power spectrum of the ECL output recorded using an optical spectrum analyzer at currents of 50 mA (i), 120 mA (ii), 180 mA (iii), and 200 mA (iv). Under certain conditions, multimode lasing is observed, although this may be eliminated via thermal tuning of one of the Vernier ring resonators. A side-mode suppression ratio of >50 dB is observed during single-mode operation.

Figure 2a plots the measured optical power at the ECL fiber-coupled output and corresponding gain chip voltage, as a function of pump current. The laser exhibits a threshold current of approximately 60 mA and achieves a maximum output power of ~25 mW at a pump current of 215 mA. To analyze laser spectra above and below threshold, we utilize an HP 70004A optical spectrum analyzer (Fig. 2c.i-iv). These measurements were performed without any adjustments to the ring resonators or the phase shifter. Just below threshold at a current of 50 mA (Fig. 2c.i), we measure the amplified spontaneous emission (ASE). Above threshold (Fig. 2c. ii), spectral narrowing is observed, with a side-mode suppression ratio (SMSR) exceeding 50 dB. Above currents of ~ 170 mA, multimode lasing behavior is observed (Fig. 2c.iii), resulting from multiple immediately adjacent longitudinal modes of the Vernier ring filter having sufficient transmission to reach threshold. This issue can be mitigated by voltage fine-tuning of the integrated heaters and may be further reduced by future modifications of the ring design to reduce the FSR of the Vernier envelope and increase side mode suppression.

To experimentally characterize the phase noise performance (and linewidth) of our ECL, we utilize an optical frequency discriminator based on an imbalanced Mach-Zehnder fiber interferometer [37-39], with varied fiber delay lengths of 4-105 m used to eliminate nulls in the discriminator sensitivity function. Measured single sideband phase noise data at a pump current of $I_p$ = 120 mA is plotted in Figure 2(b). At low frequencies, the phase noise spectrum is dominated by environmental (e.g. mechanical) noise, while above offset frequencies >$10^5$ Hz, the measured noise level closely matches that of an ideal (white frequency noise limited) laser with an intrinsic linewidth of ~65 Hz. Repeated measurements with varied pump currents and different device designs showed similar results, indicating that this noise floor is likely technical in nature (i.e., vibrations, thermal noise, drive current noise, and/or carrier noise contributions in the gain chip). Schawlow-Townes linewidth calculations according to Ref. [22] suggest that the intrinsic linewidth based on system parameters is around two orders of magnitude lower. As a result, improved performance may be possible based on the same photonic platform with further engineering work to identify and address all sources of relevant technical noise.

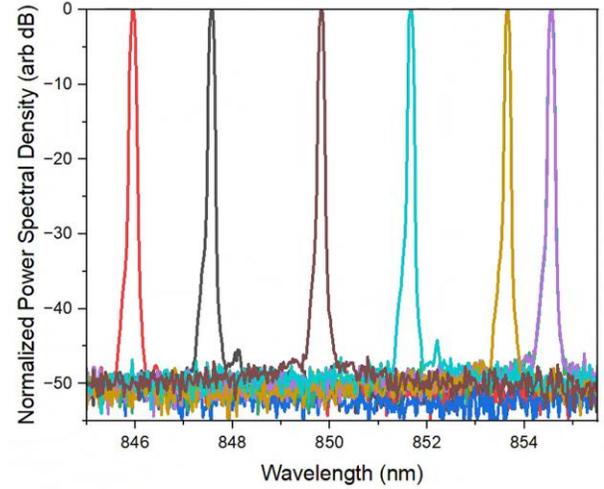

**Figure 3**. Optical spectra of the laser output as integrated heaters are used to tune the operation wavelength. Single-mode operation is repeatably achieved across a span of around ~15 nm, roughly corresponding to the RSOA gain bandwidth.

The same laser design permits tuning of the lasing wavelength across the entire gain bandwidth using the integrated heaters to adjust the refractive index of a single ring and the phase shifter region. Simultaneous tuning of one or both rings and the phase section enables continuous tuning of the cavity modes, thereby altering the lasing frequency continuously. Utilizing the phase shifters on both the ring resonators and the Sagnac loop mirror, we manually adjusted the heater powers from ~0-500 mW to measure the entire spectrum where the ECL demonstrates single-mode behavior; this relatively high heater power is due to the unoptimized heater design and low thermo-optic coefficient of $Si_3N_4$. Figure 3 illustrates a sequence of normalized laser spectra captured at seven different wavelengths during this tuning procedure. The tuning was conducted at a constant pump current of 120 mA. Initially, we adjusted the top ring resonator and observe the spectral shift. Subsequent fine-tuning of the phase shifter facilitated optimization of alignment and optical output powers. This iterative process was also applied to the phase shifter and Sagnac loop mirror heaters to achieve a uniform tuning profile. As illustrated, a spectral coverage of approximately 15 nm was achieved, and the tuning process was ultimately found to be reproducible, indicating the stability and reliability of the laser system across the specified wavelength range.

**Discussion.** In this paper, we report a $Si_3N_4$ integrated sub-100 Hz fundamental linewidth ECL at 852 nm with high side mode suppression ratio and high output power utilizing a low loss $Si_3N_4$ platform. The side mode suppression ratio compares to existing commercial device designs based in silicon nitride [40]. These enhanced performance characteristics will provide benefits to future cold atom-based systems where the coherence of the source can directly influence measurement precision. [41].

To further enhance the stability and performance of the laser design, several modifications and extensions are currently under study. First, variations to the parameters of the Vernier filters can significantly improve the suppression of nearby ring modes, permitting consistent operation at higher currents, and hence higher output powers, while maintaining single-mode operation.

Incorporating advanced phase tuning mechanisms, such as piezoelectric stress-optic tuning, would provide faster non-thermal tunability with lower power consumption and higher bandwidth. [42,43]. This improvement would facilitate high-bandwidth frequency locking of the laser to specific atomic transitions. Ruggedized packaging solutions for the dual-chip ECL system will be critical for adaptation of this device technology into practical field-deployable instruments [44,45]. Looking forward, optimized packaging must maintain stable operation and alignment while protecting the device from environmental factors like temperature fluctuations, humidity, and mechanical shocks under practical operation conditions. At the same time, output coupling to fiber or free space must be tailored to interface with the rest of a laser system, such as for the target application of cold atom-based instruments. [46,47].

In summary, we have implemented a $Si_3N_4$/GaAs-based external cavity laser around 852 nm wavelengths for future atom physics-enabled applications. The ECL achieved a free-space output power of up to 24 mW, side-mode suppression ratio greater than 50 dB, and an intrinsic linewidth of ~65 Hz. Leveraging an ultra-low loss silicon nitride based integrated circuit, we extend the cavity's roundtrip optical length to 6.5 cm, combined with dual-ring Vernier filtering to achieve tunable single-mode operation under typical conditions, within a gain bandwidth region of approximately 15 nm. This level of performance may find use in applications ranging from cold atom metrology, sensing, and LiDAR. Looking forward, these results can be readily extended to even shorter wavelengths, such as those required for i.e., rubidium (Rb), strontium (Sr), and others, within the $Si_3N_4$ platform's transparency window and where suitable gain chips are available.

**Acknowledgments.** We thank JPL's cleanroom staff, especially Dr. Daniel Wilson, Richard Muller, and Matthew Dickie for their assistance in the preparation of samples. The research was carried out at the Jet Propulsion Laboratory, California Institute of Technology, under a contract with the National Aeronautics and Space Administration (80NM0018D0004).

**Disclosures.** The authors declare no conflict of interest.

**Data availability.** Data underlying the results presented in this paper are not publicly available at this time but may be obtained from the authors upon reasonable request.

## References

1. J.C. Saywell, et al., Nature Communications, 14(1), p.7626 (2023).
2. M.J. Wright, L. Anastassiou, C. Mishra, Frontiers in Physics, 10, p.994459 (2022).
3. M.D. Lachmann, et al., Nature Communications, 12(1), p.1317 (2021).
4. A.J. McCulloch, et al., Nature communications, 4(1), p.1692 (2013).
5. S. Bize, et al., Journal of Physics B: Atomic, molecular and optical physics, 38(9), p.S449 (2005).
6. G.K. Campbell, and W.D. Phillips, Philosophical Transactions of the Royal Society A: 369(1953), pp.4078-4089 (2011).
7. B. Meyer-Hoppe, et al., Review of Scientific Instruments, 94(7) (2023).
8. O. Morizot, D. Bellair, J. De Lapeyre, et al., arXiv preprint arXiv:0704.1974 (2007).
9. N. Pie, et al., Journal of Geophysical Research: Solid Earth, 126(12), p. e2021JB022392 (2021).
10. B. Stray, et al., Nature, 602(7898), pp.590-594 (2022).
11. N. Yu, J. Kohel, J. Kellogg, and L. Maleki, Appl. Phys. B 84, 647–652 (2006).
12. A. Peters, K. Y. Chung, and S. Chu, Nature 400, 849–852 (1999).
13. V. Ménoret, P. Vermeulen, N. Le Moigne, et al., Sci. reports 8, 12300 (2018).
14. J. M. Mcguirk, G. Foster, J. Fixler, et al., Phys. Rev. A 65, 033608 (2002).
15. E. R. Elliott, et al., npj Microgravity 4, 16 (2018).
16. A. Isichenko, et al., In OFC Conference, pp. W3D-2. Optica Publishing Group, (2024).
17. N. Chauhan, et al., Nature Communications 12(1): 4685, (2021).
18. A. Isichenko, et al., 10th IEEE International Symposium on Inertial Sensors & Systems. Kauai, HI, (2023).
19. F. Lienhart, et al., Applied Physics B, 89, pp.177-180, (2007).
20. T. Komljenovic, et al., IEEE Journal of Selected Topics in Quantum Electronics, 21(6), pp.214-222 (2015).
21. I. Ghannam, et al., IEEE Journal of Selected Topics in Quantum Electronics, vol. 28, no. 1: pp. 1-10, (2022).
22. F. Youwen, v.R. Albert, J.M. Peter, et al., Opt. Express 28, 21713-21728 (2020).
23. Ultra-narrow linewidth tunable lasers. Briefings Lionix International. (n.d.).https://www.lionix-international.com/wp-content/uploads/2022/01/Briefings-Ultra-Narrow-Linewidth-Tunable-Laser_three-wavelegths-with-specification.pdf
24. C. Chen, F. Wei, X. Han, et al., Opt. Express 31, 26078-26091 (2023).
25. D.A.S. Heim,Technical Digest Series, paper STh5C.7, CLEO (2024).
26. M.A. Tran, D. Huang, J. Guo, et al., IEEE Journal of Selected Topics in Quantum Electronics", 26(2), pp.1-14 (2019).
27. N. Chauhan, J. Wang, et al., in Conference on Lasers and Electro-Optics, OSA Technical Digest, paper STh1J.2 (2020).
28. N. Chauhan, A. Isichenko, et al., Nature communications, 12(1), p.4685 (2021).
29. N. Chauhan, J. Wang, J., et al., Optics Express 30(5): 6960-6969, (2022).
30. H. Nejadriahi, and E. Kittlaus, In Metamaterials, Metadevices, and Metasystems 2023 (Vol. 12646, pp. 18-22). SPIE (2023).
31. Q. Wilmart, H. El Dirani, N. Tyler, et al., Applied Sciences, 9(2), p.255 (2019).
32. M.A. Porcel, et al., Optics & Laser Technology, 112, pp.299-306 (2019).
33. S. Tan, H. Deng, K.E. Urbanek, Optics express, 28(8), pp.12475-12486 (2020).
34. H. Dupont, (Doctoral dissertation, Tese de Mestrado, Ecole Polytechnique Fédérale de Lausanne, Lausanne), (2019).
35. Y. Zhang, et al., Opt. Express 22, 17872-17879 (2014).
36. G. Sun, et al., Optical Fiber Technology, 16(2), pp.86-89 (2010).
37. Hewlett-Packard. Phase noise characterization of microwave oscillators, frequency discriminator method. HP Product Note 11729C-2 (1985).
38. O. Llopis, P. H. Merrer, et al., Opt. Lett. 36, 2713-2715 (2011)
39. E. A. Kittlaus, et al., Nature Communications, 12(1), 4397 (2021).
40. B. Mroziewicz, Opto-Electronics Review, 16, pp.347-366 (2008).
41. A. Wicht, ISLC, (pp. 1-2). IEEE, (2018).
42. W. Jin, R. G. Polcawich, et al., Opt. Express 26, 3174-3187, (2018).
43. J. Wang, Opt. Express 30, 31816-31827 (2022).
44. M. Alalusi, et al., In Fiber Optic Sensors and Applications VI (Vol. 7316, pp. 235-247). SPIE (2009).
45. J.P. McGilligan, et al., Review of Scientific Instruments, 93(9), (2022).
46. C. Ropp, et al., Light: Science & Applications, 12(1), p.83 (2023).
47. A. Isichenko, Nature communications, 14(1), p.3080, (2023).